# Channel State Prediction, Feedback and Scheduling for a Multiuser MIMO-OFDM Downlink

Hooman Shirani-Mehr
shiranim@usc.edu

Daniel N. Liu
danielnl@usc.edu

Giuseppe Caire
caire@usc.edu

*Abstract*— We consider the downlink of a MIMO-OFDM wireless systems where the base-station (BS) has $M$ antennas and serves $K \geq M$ single-antenna user terminals (UTs). Users estimate their channel vectors from common downlink pilot symbols and feed back a prediction, which is used by the BS to compute the linear beamforming matrix for the next time slot and to select the users to be served according to the proportional fair scheduling (PFS) algorithm. We consider a realistic physical channel model used as a benchmark in standardization and some alternatives for the channel estimation and prediction scheme. We show that a parametric method based on ESPRIT is able to accurately predict the channel even for relatively high user mobility. However, there exists a class of channels characterized by large Doppler spread (high mobility) and clustered angular spread for which prediction is intrinsically difficult and all considered methods fail. We propose a modified PFS that take into account the "predictability" state of the UTs, and significantly outperform the classical PFS in the presence of prediction errors. The main conclusion of this work is that multiuser MIMO downlink yields very good performance even in the presence of high mobility users, provided that the non-predictable users are handled appropriately.[1]

## I. INTRODUCTION

Multiuser MIMO downlink schemes are expected to become the corner-stone of future wireless cellular systems since they can achieve a downlink capacity that scales linearly with the number of base station (BS) transmit antennas $M$, even though each user terminal (UT) has only a single antenna. In order to accomplish this linear capacity scaling with $M$, accurate Channel State Information (CSIT) needs to be provided to the BS [6], [10], [5]. Following the current standardization trend of systems like 3GPP LTE and IEEE 802.16m (mobile WiMax) [2] and [1], we assume that the BS broadcasts downlink pilot symbols in order to allow each UT to estimate its own channel state vector. We assume a frequency division duplexing (FDD) scheme, where CSIT feedback must be implemented by explicit closed-loop feedback. Since fading channels are generally time-varying, the UTs must *predict* their channel state and feed back this prediction via some CSIT feedback channel (uplink). In this paper we assume an ideal unquantized feedback and focus uniquely on the effects of channel estimation and prediction: of course, any form of explicit CSIT feedback can only decrease the CSIT accuracy with respect to pure prediction and ideal feedback.

[1]All authors are with the EE Department of the University of Southern California, Los Angeles, CA. This work was partially supported by the Collaborative USC-ETRI Project 53-4503-0781.

The BS computes the downlink beamforming matrix based on the predicted user channels. Since in general we have $K \geq M$ users, the BS selects a subset of $S \leq M$ users that are served at each point in time. This selection and corresponding rate allocation is known as *scheduling*. In this paper we consider the widely used *proportional fair scheduling* (PFS) algorithm [12], [11].

Observing the performance of two rather well-known channel prediction methods applied to the SCM channel model [3], which is used as a benchmark in standardization, we notice that channels with well separated angles of arrivals and/or very low maximum Doppler spread are easy to predict. In contrast, there exist a class of channels with high Doppler spread and clustered angles of arrival for which all prediction method fail. A more refined analysis based on the Cramer-Rao Bound [7], not reported here because of lack of space, shows that the "non-predictability" of these channels is an intrinsic property and not just a shortcoming of the algorithms considered. A naive application of the PFS scheme for these *non-predictable* users yields large performance degradation, because scheduling computed from inaccurate CSIT. We propose a simple modification of the basic PSF algorithm that is both very simple and provides significant improvements.

For the application of the proposed modified PFS (MPFS) scheme it is essential that each UT informs the BS whether its channel is predictable or non-predictable. Fig. 1 illustrates a block diagram of the process run at each UTs: the UT keeps predicting its channel from the downlink pilot symbols. Also, it compares its prediction with the actual channel in the next time slot, and keeps track of the prediction error. The UT notifies its predictability status to the BS by comparing the estimated prediction error with a suitable threshold. UTs change state (from predictable to non-predictable and vice versa) at a rate much slower than the CSIT feedback rate. Notice also that the non-predictable users need not feeding back their CSIT prediction. Therefore, as a by-product, this scheme lowers the requirement of the CSIT feedback channel by avoiding to feedback inaccurate and useless information.

## II. SYSTEM MODEL

The MIMO-OFDM downlink model with $M$ BS antennas, one antenna at each UT and $N$ subcarriers is given by

$$y_k[t,n] = \mathbf{h}_k^{\mathsf{H}}[t,n]\mathbf{x}[t,n] + z_k[t,n] \qquad (1)$$

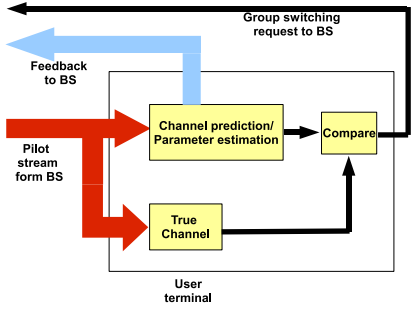

Fig. 1. Block diagram of the prediction and prediction error tracking process at each UT.

for $k = 1, \ldots, K$, and $n = 1, \ldots, N$, where $t$ denotes discrete time (that ticks at the OFDM symbol rate), $n$ is the subcarrier index, $k$ is the UT index, $y_k[t,n]$ is the received signal at UT $k$, time $t$ and subcarrier $n$, $\mathbf{h}_k[t,n]$ is the corresponding $M \times 1$ channel vector, $\mathbf{x}[t,n]$ is the transmit signal vector and $z_k[t,n] \sim \mathcal{CN}(0, N_0)$ is AWGN. The channel vector $\mathbf{h}_k[t,n]$ has (frequency-domain) channel coefficients $H_{k,m}[t,n]$ for BS antenna index $m = 1, \ldots, M$. We consider the SCM channel model used for urban micro-cell, urban macro-cell and suburban macro-cell fading environments [3]. This model considers $P$ clusters of scatterers where each cluster corresponds to a path (same relative delay) and each path consists of $R_p$ subpaths. Each subpath is characterized by an angle of arrival and by a complex amplitude coefficient. After standard manipulation, the frequency-domain channel (dropping indices $k, m$ by symmetry and for the sake of notation simplicity) can be written in the form of

$$H[t,n] = \sum_{p=1}^{P} \sum_{r=1}^{R_p} A_{r,p} e^{-j2\pi \tau_p n} e^{j2\pi \zeta_{r,p} t} \quad (2)$$

where $A_{r,p}$ is the complex amplitude coefficient of the $(r,p)$-th subpath, $\zeta_{r,p}$ is the Doppler frequency shift of the $(r,p)$-th subpath (multiplies by the OFDM symbol duration) and $\tau_p$ is the delay of the $p$-th path (multiplied by the subcarrier spacing).

The BS makes use of linear zero-forcing beamforming (ZFBF), i.e., it produces the the transmitted signal vector as $\mathbf{x}[t,n] = \mathbf{V}[t,n]\mathbf{u}[t,n]$ where $\mathbf{u}[t,n]$ is a vector of user coded symbols and $\mathbf{V}[t,n]$ is the beamforming matrix. At each time-frequency slot $[t,n]$, a subset of users $\mathcal{S}[t,n]$ of size $S[t,n] \leq M$ is selected according the scheduling algorithm. Therefore, $\mathbf{V}[t,n]$ has dimensions $M \times S[t,n]$. For a given set of selected users $\mathcal{S}[t,n]$ of size $S[t,n]$, the corresponding ZFBF matrix is obtained by first computing the pseudoinverse of the matrix $\widehat{\mathbf{H}}[t,n]$ with columns $\{\widehat{\mathbf{h}}_k[t,n] : k \in \mathcal{S}[t,n]\}$, where $\widehat{\mathbf{h}}[t,n]$ indicates the prediction of the actual channel $\mathbf{h}_k[t,n]$ available at the BS, and then normalizing the pseudoinverse columns to unit norm such that the average transmit power per subcarrier is $\mathbb{E}[|\mathbf{x}[t,n]|^2] = P$. Letting $\widehat{\mathbf{v}}_k[t,n]$ denote the beamforming vector for user $k \in \mathcal{S}[t,n]$, and $p_k[t,n]$ its allocated power,

the *nominal* achievable rate of user $k$ is given by

$$\widetilde{R}_k[t,n] = \log\left(1 + |\widehat{\mathbf{h}}_k^{\mathsf{H}}[t,n]\widehat{\mathbf{v}}_k[t,n]|^2 p_k[t,n]\right) \quad (3)$$

We refer to this rate as "nominal" since this is what the BS assumes, based on the available CSIT in the current slot. In the case of non-perfect CSIT, user $k$ suffers from multiuser interference since ZFBF is generally mismatched. Here we make the *optimistic* assumption that, no matter how mismatched the beamforming matrix is, an instantaneous rate equal to

$$R_k[t,n] = \log\left(1 + \frac{\left|\mathbf{h}_k^{\mathsf{H}}[t,n]\widehat{\mathbf{v}}_k[t,n]\right|^2 p_k[t,n]}{1 + \sum_{j \neq k} \left|\mathbf{h}_k^{\mathsf{H}}[t,n]\widehat{\mathbf{v}}_j[t,n]\right|^2 p_j[t,n]}\right) \quad (4)$$

is actually achieved. In practice, some mechanism for fast rate adaptation that cope with the BS mismatched CSIT can be used in order to approach the actual rate. Such optimistic achievability assumption provides an upper bound to the performance with any form of rate adaptation, e.g., as obtained by incremental redundancy ARQ and the use of rateless codes [15].

## III. SCHEDULING

If the CSIT of user $k$ is accurate, i.e., if $\widehat{\mathbf{h}}_k \approx \mathbf{h}_k$, then $\widetilde{R}_k[t,n] \approx R_k[t,n]$. On the other hand, if the channel of user $k$ cannot be accurately predicted, then $R_k[t,n] \ll \widetilde{R}_k[t,n]$ since the multiuser interference in the denominator of the SINR in (4) is large. The PFS algorithm selects at each $[t,n]$ a user subset $\mathcal{S}[t,n]$ in order to maximize the *weighted rate sum* $\sum_k \frac{\widetilde{R}_k[t,n]}{T_k[t]}$ where $\{\widetilde{R}_k[t,n]\}$ are the nominal rates at time-frequency slot $[t,n]$ and $\{T_k[t]\}$ are the actually achieved long-term throughputs at time $t$, which are recursively updated according to the rule

$$T_k[t] = (1-\beta)T_k[t-1] + \beta \frac{1}{N} \sum_{n=1}^{N} R_k[t,n] \quad (5)$$

where $\beta \in (0,1)$ is the factor that controls the average delay. This selection can be performed, for example, by using the greedy user selection algorithm of [8]. Consider a "non-predictable" user $k$. For what observed above, since $T_k[t]$ is updated by using the actually achievable rate, it is likely that $T_k[t]$ will become very small. Hence, user $k$ will have a large weight in the PFS objective function and will be served very frequently. It follows that non-predictable users "eat-up" a lot of transmission resources from the system. Unfortunately, since they suffer from high multiuser interference power, their rate will remain low. Hence, due to scheduling, the presence of non-predictable users impacts negatively also the performance of predictable users.

In order to avoid this problem, we propose the MPFS scheme where users are partitioned into a predictable set of size $K_p$ and a non-predictable set of size $K_{np} = K - K_p$. PFS is applied with the additional constraint that if $k \in K_{np}$ is selected on slot $[t,n]$, no other user can be selected on the same time-frequency slot. Furthermore, since users in the non-predictable class have unreliable CSIT, the BS serves them using standard transmit diversity (i.e., it does not try to beamform). The nominal rate for these users is given by $\widetilde{R}_k[t,n] = \log\left(1 + \frac{P}{M}|\widehat{\mathbf{h}}_k[t,n]|^2\right)$. Notice that in this case only a coarse form of CSIT is needed, which in practice amounts to knowing the user instantaneous SNR with some degree of accuracy.

For the sake of comparison, we consider also a "hard-fairness" scheduling (HFS) scheme that aims at maximizing the minimum throughput over all users, i.e., HFS solves $\max \min_k T_k$. Predictable users and non-predictable users are served in separated time-frequency slots. Given the symmetry of the channel with respect to all frequencies, we can focus on one subcarrier and let $\alpha_p$ and $\alpha_{np} = 1 - \alpha_p$ denote the fraction of time slots allocated to predictable and non-predictable users. Non-predictable users are served in round-robin using transmit diversity. Modeling the user channel square magnitude $z = |\mathbf{h}_k|^2$ as a Chi-square RV with $2M$ degrees of freedom, the individual average throughput of any of these users, by symmetry, is given by

$$T_{np} = \frac{1}{K_{np}} \mathbb{E}\left[\log\left(1 + \frac{\mathcal{P}}{M}|\mathbf{h}_k|^2\right)\right]$$
$$= \frac{1}{K_{np}}\left(\frac{M}{\mathcal{P}}\right)^M \frac{1}{(M-1)!} \mathcal{I}_M\left(\frac{M}{\mathcal{P}}\right) \quad (6)$$

where $\mathcal{I}_n(\mu) = \int_0^\infty t^{n-1}\ln(1+t)e^{-\mu t}dt = (n-1)!e^\mu \sum_{k=1}^n \frac{\Gamma(-n+k,\mu)}{\mu^k}$ [4], where $\Gamma(.,.)$ is the complementary incomplete gamma function defined by [9] $\Gamma(\alpha,x) = \int_x^\infty t^{\alpha-1}e^{-t}dt$. The predictable users can be served using regular PFS applied in their fraction of allocated slots. The average rate per user can be lower bounded by considering equal power allocation as

$$\frac{1}{K_p}\sum_{k=1}^M \mathbb{E}\left[\log\left(1 + \frac{\mathcal{P}}{M}|\mathbf{h}_k^H \mathbf{v}_k|^2\right)\right] \leq T_p \quad (7)$$

For i.i.d. Rayleigh fading and using the fact that $|\mathbf{h}_k^H \mathbf{v}_k|^2$ is Chi-squared with 2 degrees of freedom, the LHS of (7) takes on the familiar form

$$\frac{M}{K_p} e^{\frac{M}{\mathcal{P}}} E_i\left(1, \frac{M}{\mathcal{P}}\right) \quad (8)$$

where $E_i(n,x) = \int_1^\infty \frac{e^{-xt}}{t^n}dt$.

The max-min solution is achieved by balancing the throughputs. By letting $\alpha_p T_p = (1-\alpha_p)T_{np}$ we find $\alpha_p = \frac{T_{np}}{T_p+T_{np}}$. In practice, since the statistics of the channels may be non-Rayleigh and the application of PFS for the predictable users may improve $T_p$ with respect to the above lower bound, an adaptive slot allocation should take care of updating the value of $\alpha_p$ in the HFS scheme. Our simulations based on the SCM channel model show that dimensioning $\alpha_p$ according to the above rule of thumb yields values very close to the actual rate-balancing state.

## IV. CHANNEL PREDICTION SCHEMES

We consider the downlink common pilot arrangement of Fig. 2. Pilots are separated by $D_t$ OFDM symbols in time and by $D_f$ subcarriers in frequency. A total of $N_t \times N_f$ pilots per antenna are collected over a window of duration $N_t D_t$, and used for prediction. Pilots for different transmit antennas are mutually orthogonal. In order to avoid aliasing, we need that $D_f \leq \frac{1}{\tau_{\max}}$ and $D_t \leq \frac{1}{2\zeta_{\max}}$ where $\tau_{\max}$ is the maximum delay and $\zeta_{\max}$ is the maximum Doppler frequency shift.

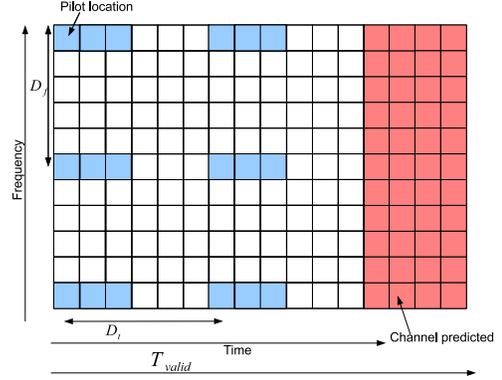

Fig. 2. Pilot pattern.

By defining $\omega_{r,p} = 2\pi\zeta_{r,p}D_t$, $v_p = 2\pi\tau_p D_f$, the output of the pilot observation (up to an irrelevant phase shift due to the location of the pilot subcarriers) can be written in the form of a 2-dimensional sum of sinusoids in additive white Gaussian noise as

$$\widehat{H}[qD_t, mD_f] = \sum_{p=1}^P \sum_{r=1}^{R_p} A_{r,p} e^{j(\omega_{r,p}l + v_p m)} + z[l,m] \quad (9)$$

Channel prediction under this model reduces to estimating the parameters of the sinusoidal components. We follow the method of [17], based on the repeated application of ESPRIT (details are not included due to lack of space). In [7] we have also considered non-linear Least-Squares, and various approximations of ML estimation [13], [19], [16], and we concluded that, for this problem, the ESPRIT method offers the best tradeoff between complexity and performance.

An alternative channel prediction scheme consists of modeling the discrete-time (time-domain) channel impulse response as Gaussian stationary vector process and using MMSE linear prediction (Wiener filter) [18]. The time-domain channel impulse response $(h(t,0),\ldots,h(t,L-1))$ is related to the frequency-domain channel by the DFT

$$H[t,n] = \sum_{l=0}^{L-1} h(t,l)e^{-j2\pi ln/N} \quad (10)$$

Since the channel is "sampled" by the downlink pilot grid every $D_t$ OFDM symbols, it follows that prediction can be done in steps of $D_t$ symbols. Then, the BS treats the channel as piecewise constant for blocks of length $D_t$. For simplicity, we assume that the time-domain channel impulse response is measured by sending a single impulse in time followed by $N_f - 1$ zeros, with $N_f \geq L$, with impulse power $N_f P$. In this way, the frequency domain pilot grid of Fig. 2 and the time-domain pilot scheme have the same overhead in terms of pilot symbols insertion ratio and total pilot power. Both the ESPRIT-based and the Wiener-based schemes need to compute and update the sample covariance matrix of the received signal. This reduces to a recursive computation that is very similar to standard RLS.

It is interesting to notice that, in terms of system design, the two prediction methods yield to quite different CSIT feedback schemes. Wiener prediction produces a new predicted channel vector every $D_t$ OFDM symbols, and this needs to be fed back within this delay. In contrast, ESPRIT produces a set of channel parameters that is used to extrapolate the channel over a large block (see [17] for a quantitative estimate of the duration over which the channel parameters can be considered to be constant ). The UT will estimate the channel parameters $\{A_{r,p}, \omega_{r,p}, v_p\}$ from a block of pilots of duration $D_t N_t$ OFDM symbols, extrapolate the channel for the next block, and feed back any convenient quantized representation of the extrapolated trajectory.[2] Fig 3 illustrates the different feedback modes suited to these prediction methods. As said, the feedback for ESPRIT is "batch" while for Wiener it is a continuous low-rate transmission.

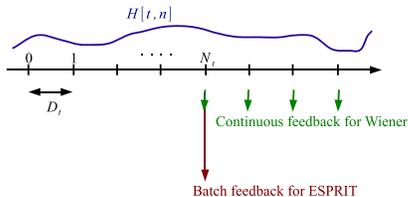

Fig. 3. Batch versus frame by frame feedback.

## V. SIMULATIONS AND DISCUSSION

We consider a system with parameters given in Table I.

Maximum Doppler frequency shift is given by $\zeta_{max} = \frac{f_c v}{c}$ where $c$ is the speed of light, $v$ is the speed of user and $f_c$ is the carrier frequency. Duration of validity of parameters is given by $T_{\text{valid}} = \sqrt{\frac{c r_{\min}}{3 f_c v^2}}$ where $r_{\min}$ is the distance from user to nearest scatterer and we assume that $r_{\min} = 600$m. The corresponding values are given in Table II.

We have compared the two prediction methods by considering 4 different scenarios. Our findings are summarized by Table III. When Doppler frequencies are well-separated, the Doppler frequency components of the channel can be accurately estimated by ESPRIT for both high speed and low speed user. In contrast, when Doppler frequencies are packed, if user is low speed then channel varies slowly and can be predicted by Wiener or even treated as piecewise constant while for high speed users both prediction methods fail. This class of channels can be considered as "non-predictable", i.e., the accuracy of prediction is not sufficient for the purpose of MIMO downlink beamforming.

[2] In general, feeding back the channel coefficient is not the optimal strategy for a given target CSIT distortion. A rate-distortion approach to channel state feedback when the channel is modeled as a correlated Gaussian source is proposed in [14]. In the case of the SCM channel model, we leave this interesting rate-distortion problem for future work.

TABLE I
SYSTEM SPECIFICATIONS

| Description | Value |
|---|---|
| $1/T$, Sampling frequency | 3.84MHz |
| $f_c$, Carrier frequency | 2.6GHz |
| $N$, Number of subcarriers | 256 |
| $N_a$, Number of active subcarriers | 200 |
| Length of CP | 64 |
| $\Delta f$, Subcarrier frequency spacing | 15KHz |
| $T_{sym}$, OFDM symbol with CP | 83.33$\mu$s |
| $N_f$, Number of Pilot subcarriers | 50 |
| $N_t$, Number of Pilot OFDM symbols | 100 |
| $D_f$, Pilot subcarrier spacing | 4 |
| $D_t$, Pilot OFDM symbol spacing | 20 |
| $\tau_{max}$, Maximum delay | 16.67$\mu$s |

TABLE II
MAXIMUM DOPPLER FREQUENCY AND DURATION OF VALIDITY OF MODEL FOR UT WITH DIFFERENT SPEEDS

| speed | $v$ (Km/h) | $\zeta_{max}$ (Hz) | $T_{\text{valid}}$ (s) | $T_{\text{valid}}$ (OFDM symbols) |
|---|---|---|---|---|
| High | 75 | 180.56 | 0.231 | 2700 |
| Low | 5 | 12.04 | 3.458 | 41500 |

Next, we consider a BS with $M = 4$ transmit antennas, $K = 8$ single antenna UTs with $v = 75$km/h (high mobility), SCM channel model and SNR = 20dB. $K_{np} = 2$ users (users 1 and 2) have packed Doppler frequencies and their prediction is very poor for both ESPRIT and Wiener prediction and $K_p = 6$ users (users 3-8) have well-separated Doppler frequencies and their channels are predicted reliably by applying ESPRIT. Figs. 4, 5 and 6 and Table IV demonstrate simulation results for different scheduling methods. Looking at the activity fractions of Table IV, i.e., the fraction of slots in which a given user is scheduled, we notice that the naive PFS that makes use of unreliable CSIT tends to serve the non-predictable users (users 1 and 2) very often. This confirm the behavior that we described qualitatively in Section III. This problem is eliminated by the proposed MPFS scheme.

We observe that the proposed MPFS scheme achieve the best performance both in terms of throughput sum $\sum_k T_k$, and in terms of sum of log of throughputs $\sum_k \log T_k$ (proportional fairness objective function [12], [11]).

We conclude that by treating predictable and non-predictable users in two classes as in the proposed MPFS scheme, and by using the appropriate parametric prediction scheme, such as the one based on ESPRIT, multiuser MIMO downlink can be successfully applied also to users with relatively high mobility.

TABLE III
BEST PREDICTION METHOD

|  | Well-separated Dopplers | Packed Dopplers |
|---|---|---|
| Low speed user | ESPRIT | Wiener |
| High speed user | ESPRIT | All fail |

TABLE IV
ACTIVITY FRACTIONS

|  | 1 | 2 | 3 | 4 | 5 | 6 | 7 | 8 |
|---|---|---|---|---|---|---|---|---|
| PFS | 0.74 | 0.79 | 0.35 | 0.37 | 0.36 | 0.35 | 0.36 | 0.42 |
| MPFS | 0.14 | 0.15 | 0.44 | 0.44 | 0.44 | 0.44 | 0.44 | 0.53 |
| HFS | 0.26 | 0.26 | 0.29 | 0.29 | 0.29 | 0.29 | 0.29 | 0.35 |

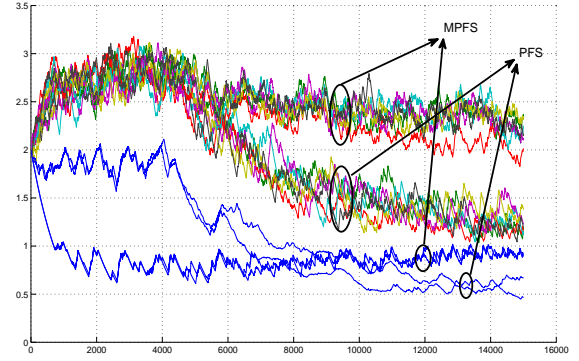

Fig. 4. Evolution of $T_k[t]$ over time for different users.

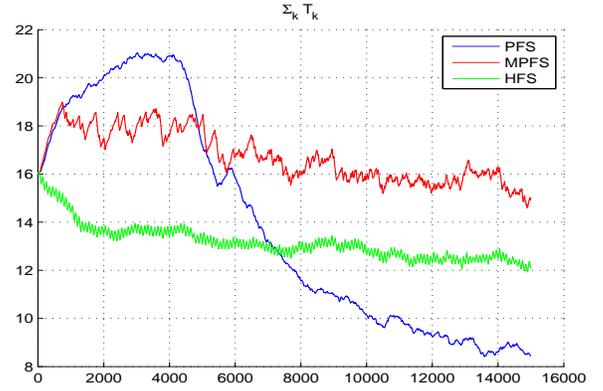

Fig. 5. Sum rate for PFS, MPFS and HFS for the configuration with 2 non-predictable and 6 predictable high-speed users.

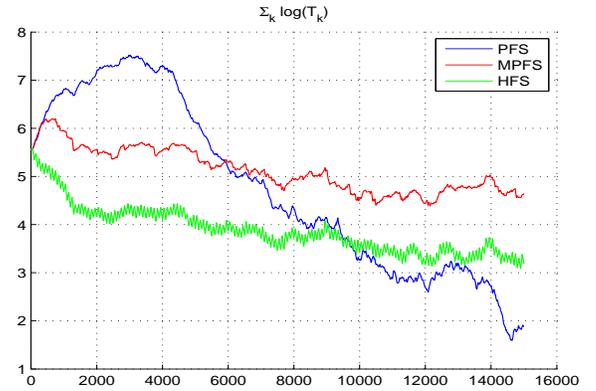

Fig. 6. Sum log rate for PFS, MPFS and HFS for the configuration with 2 non-predictable and 6 predictable high-speed users.